\documentclass[twocolumn,prl,aps,superbib,tightenlines,floatfix,superscriptaddress,showpacs]{revtex4}
\usepackage[T1]{fontenc}
\usepackage[latin9]{inputenc}
\usepackage{amsmath}
\usepackage{graphicx}
\usepackage{amssymb}

\begin{document}

\title{Dynamics of Spin Relaxation in Finite-Size 2D Systems: an Exact Solution}

\author{Valeriy A. Slipko}
\affiliation{Department of Physics and Astronomy and USC
Nanocenter, University of South Carolina, Columbia, SC 29208, USA}
\affiliation{ Department of Physics and Technology, V. N. Karazin
Kharkov National University, Kharkov 61077, Ukraine }

\author{Yuriy V. Pershin}
\email{pershin@physics.sc.edu}

\affiliation{Department of Physics and Astronomy and USC
Nanocenter, University of South Carolina, Columbia, SC 29208, USA}

\begin{abstract}
We find an exact solution for the problem of electron spin relaxation in a 2D circle with Rashba spin-orbit interaction.
 Our analysis shows that the spin relaxation in finite-size regions involves three stages and is described by multiple spin relaxation times. It is important that the longest spin relaxation time increases with decrease in system radius but always remains finite. Therefore, at long times, the spin polarization in small 2D systems decays exponentially with a size-dependent rate. This prediction is supported by results of Monte Carlo simulations.
\end{abstract}

 \maketitle

The problem of D'yakonov-Perel'~\cite{Dyakonov72a,Dyakonov86a} spin relaxation in two-dimensional (2D) systems have attracted wide attention~\cite{Dyakonov72a,Dyakonov86a,Sherman03a,Burkov04a,Saikin04a,Pershin04b,Pershin05a,Lyubinskiy06b,Pershin07a,Weng08a,Kleinert09a,pershin10a,Tokatly10a,Tokatly10b} because of its fundamental importance for the field of spintronics~\cite{Zutic04a,Wu10a}. However, the spin relaxation in systems with boundaries is even more important because boundaries are naturally present in all electronic devices.  There are only several examples in the literature where the influence of boundary conditions on D'yakonov-Perel' spin relaxation have been explored theoretically and/or experimentally.
These examples include investigations of spin relaxation in 2D channels~\cite{Kiselev00a,Schwab06a,Holleitner07a,Chang09a,Frolov10a,Liu10a}, 2D half-space~\cite{Pershin05c}, 2D systems with antidots~\cite{Pershin04a}, large quantum dots~\cite{Lyubinskiy06a,Koop08a}, and one-dimensional (1D) finite-length wires~\cite{Slipko11a}. Both available experimental and theoretical results indicate that typically in the diffusive spin transport regime the electron spin life time is longer in systems with boundaries.

In this paper we find an exact solution for the problem of electron spin relaxation in finite-size 2D systems. Specifically, we consider the dynamics of electron spin relaxation in a 2D circle made of a semiconductor structure with Rashba-type spin-orbit interaction. Naively, one may think that in small systems the spin relaxation is incomplete as the spin precession angle across the system is small. However, in such a situation, the different effect plays role: the non-commutativity of spin rotations. Because of this effect, the electron spin precession angle can largely exceed the maximum rotation angle allowed by naive geometrical considerations. To the best of our knowledge, the spin relaxation in small systems was investigated previously only in Ref. ~\onlinecite{Lyubinskiy06a}. This previous study provides only an asymptotic value of the spin relaxation time without giving details on how the whole process of spin relaxation occurs. In the present paper, we show that the spin relaxation process in finite-size systems is intrinsically complex. The exact solution of this fundamental problem involves an infinite number of spin relaxation constants. At long times, however, only the slowest decaying component survives and the spin polarization exhibit a slow size-dependent exponential decay. Our exact analytical solution is obtained using the Laplace transform and is confirmed by Monte Carlo simulations of spin dynamics. This work thus provides an important missing part of spin relaxation theory.

\begin{figure}[b]
 \begin{center}
\includegraphics[angle=0,width=6.0cm]{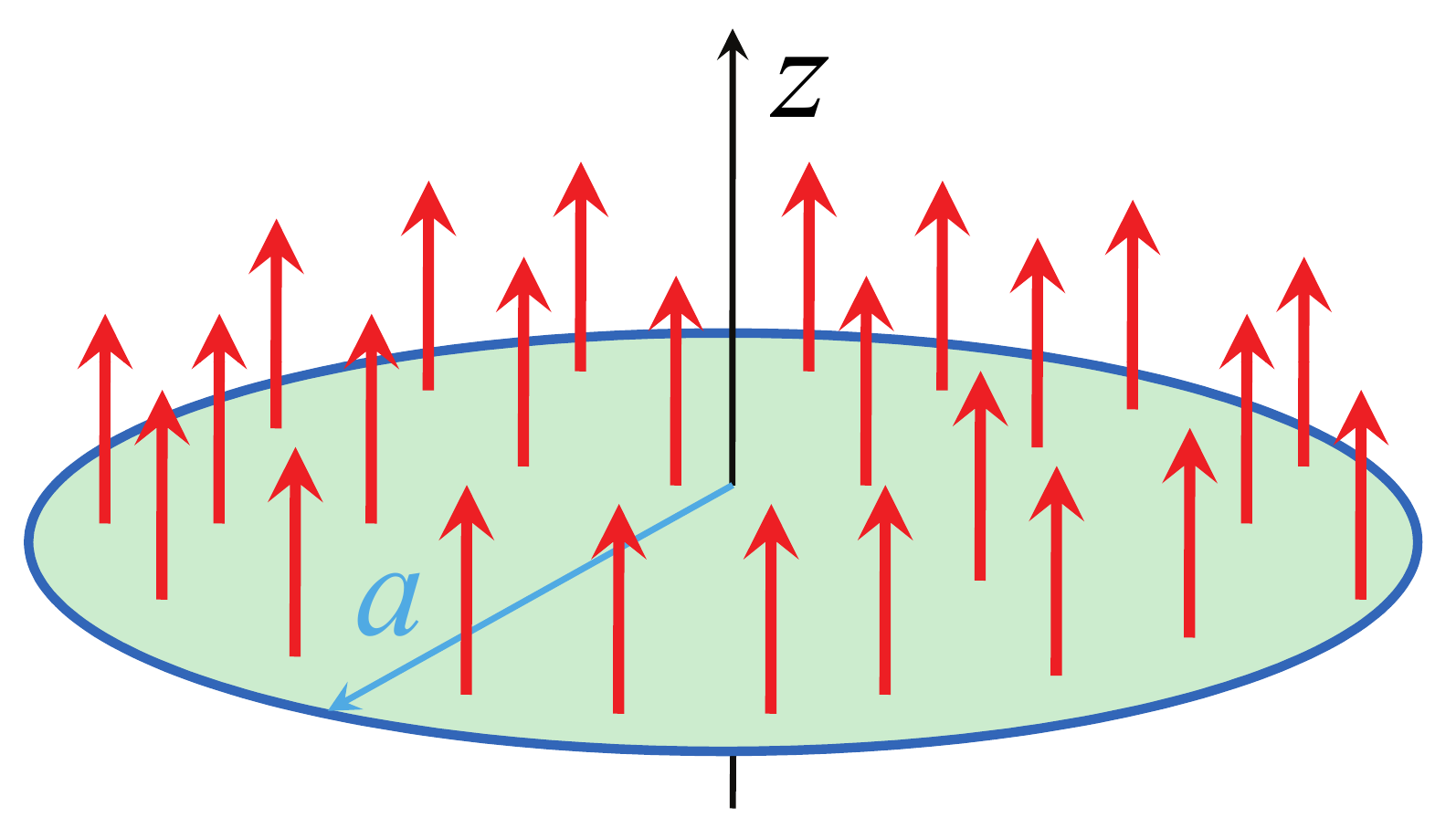}
\end{center}
 \caption{(Color online) Schematic of a circular 2D region of a radius $a$ with electron spin polarization initially pointing in $z$ direction perpendicular to the circle plane. \label{fig0}}
\end{figure}

Let us consider the dynamics of electron spin polarization in a  finite-size 2D electron system such as the circle shown in Fig. \ref{fig0}. The set of diffusion equations~\cite{Burkov04a,Galitski06a,arXiv_1007_0853v1}  for spin polarization is given by
\begin{eqnarray}
\frac{1}{D}\frac{\partial \mathbf{S}_{\textrm{in}}}{\partial t}=\Delta \mathbf{S}_{\textrm{in}}+2\eta\nabla
S_z-\eta^2 \mathbf{S}_{\textrm{in}}\label{SperpEq}, \\
\frac{1}{D}\frac{\partial S_z}{\partial t}=\Delta S_z-2\eta \nabla\cdot \mathbf{S}_{\textrm{in}}-2\eta^2 S_z
\label{SzEq},
\end{eqnarray}
where $\mathbf{S}_{\textrm{in}}$ and $S_z$ are the in-plane and $z$ components of spin polarization, respectively, $D=l^2/(2\tau )$ is the diffusion coefficient, $l$ is the mean free path, $\tau$ is the momentum relaxation time, and $\eta$ is the spin rotation angle per unit length (spin rotations are induced by the Rashba spin-orbit interaction~\cite{Bychkov84a}). The applicability limits of Eqs. (\ref{SperpEq}) and (\ref{SzEq}) are $\lambda\ll l \ll a$ and $l \ll \eta^{-1}$, where $\lambda$ is the electron's de Broglie wavelength, and $a$ is the characteristic system size (in the context of this work, $a$ is the radius of the circle). We also note that Eqs. (\ref{SperpEq}) and (\ref{SzEq}) are valid for any value of $a\eta$. Eqs. (\ref{SperpEq}) and (\ref{SzEq}) are supplemented by the standard boundary conditions~\cite{Galitski06a}
\begin{eqnarray}
\left(\frac{\partial S_n}{\partial n}+\eta S_z\right)_\Gamma=0, \left(\frac{\partial S_z}{\partial n}-\eta S_n\right)_\Gamma=0,
\left(\frac{\partial S_\Upsilon}{\partial n}\right)_\Gamma=0, \;\;
\label{BCGenTan}
\end{eqnarray}
where $\mathbf{n}$ and $\mathbf{\Upsilon}$ are the normal in-plane and tangential vectors to the boundary $\Gamma$, respectively.

Next, we would like to reduce the set of Eqs. (\ref{SperpEq}) and (\ref{SzEq}) to a single equation for $S_z$. Introducing
$u=\nabla\cdot\mathbf{S}_{\textrm{in}}$ and $v=(\nabla\times\mathbf{S}_{\textrm{in}})_z$, Eqs. (\ref{SperpEq}) and (\ref{SzEq}) can be rewritten as
\begin{eqnarray}
u_t=\Delta u+2\Delta S_z- u
\label{uEq}, \\
(S_z)_t=\Delta S_z-2S_z-2u
\label{wEq}, \\
v_t=\Delta v -v
\label{vEq}.
\end{eqnarray}
Here, $\Delta$ is the 2D Laplace operator, the time is measured in the units of $t_s=(D\eta^2)^{-1}$, and the coordinates are measured in the units of $\eta^{-1}$. Such a convention is used below if not stated otherwise. Combining Eqs. (\ref{uEq}) and (\ref{wEq}) we readily get
\begin{eqnarray}
(S_z)_{tt}-2\Delta (S_z)_t+3(S_z)_t+\Delta^2 S_z +\Delta S_z+2S_z=0.
\label{wtt}
\end{eqnarray}
The Laplace transform of Eq. (\ref{wtt}) is given by
\begin{eqnarray}
\Delta^2 \tilde{S_z}+(1-2p)\Delta \tilde{S_z}+(p^2+3p+2)\tilde{S_z} \nonumber\\
=(p+1)S_z|_{t=0}-\Delta S_z|_{t=0}-2u|_{t=0},
\label{LTwGEq}
\end{eqnarray}
where $\tilde{S_z}$ is the Laplace transform of $S_z$ to the complex $p$-domain.  In the above equation,
the time derivative of $S_z$ at $t=0$ was substituted from Eq. (\ref{wEq}).

In what follows we consider the relaxation of homogeneous spin polarization pointing in $z$ direction perpendicular to the plane of 2D circle (see Fig. \ref{fig0}).
Taking into account the axial symmetry of the problem, the initial and boundary conditions read
\begin{eqnarray}
u(r,t=0)=\frac{\partial (rS_{r}(r,t=0))}{r\partial r}=0,
S_z(r,t=0)=S_0, \label{ICS} \\
\left. \left(\frac{\partial S_r}{\partial r}+ S_z\right)\right|_{r=a}=0, \; \left. \left(\frac{\partial S_z}{\partial r}-
S_r\right)\right|_{r=a}=0.\label{BCCircle}\;\;
\end{eqnarray}
We note that the function $v=\partial (rS_{\phi})/(r\partial r)$ is safely taken out of the consideration since the solution $v(r,t)=0$ satisfies Eq. (\ref{vEq}) with the boundary condition $\partial S_{\phi} /\partial r|_{r=a}=0$ and the initial condition $v(r,0)=0$.
Applying the initial conditions (\ref{ICS}), Eq. (\ref{LTwGEq}) simplifies to
\begin{equation}
\Delta^2 \tilde{S_z}(r)+(1-2p)\Delta \tilde{S_z}(r)+(p^2+3p+2)\tilde{S_z}(r)
=(p+1)S_0.\;\;
\label{LTwEq}
\end{equation}

The general solution of Eq. (\ref{LTwEq}) can be found by the factorization of its left-hand side and be presented as
\begin{eqnarray}
\tilde{S_z}(r)=A_{1}J_0(k_{1}r)+A_{2}J_0(k_{2}r)+B_{1}N_0(k_{1}r)\nonumber \\
+B_{2}N_0(k_{2}r)+\frac{S_0}{p+2},
\label{LTwSol}
\end{eqnarray}
where $A_{1,2}, B_{1,2}$  are arbitrary constants, $J_0(z)$ and  $N_0(z)$ are the zeroth order Bessel and Neumann functions, respectively, and
\begin{eqnarray}
k^2_{1,2}=-p+\frac{1}{2}\pm 2i\sqrt{p+\frac{7}{16}}.
\label{kpm}
\end{eqnarray}
In the case of the circle, $B_{1,2}=0$ since $N_0(x)$ diverges as $x\rightarrow 0$. Moreover, the actual choice of two branches corresponding to $\pm$ in Eq. (\ref{kpm}) is not essential.
We may make a branch cut along the line $p<-7/16$, Im$(p)=0$ in the plane of complex $p$ and define two branches by the conditions
$k^2_{1,2}(p=0)=(1\pm i\sqrt{7})/2$.

Although the radial component of spin polarization equals zero at $t=0$, it becomes different than zero at $t>0$ similarly to
the case of spin relaxation in rings \cite{Slipko11b}. With a help of Laplace transform of Eq. (\ref{wEq}), we express
$\tilde{S_r}(r)$ through $\tilde{S_z}(r)$ as
\begin{eqnarray}
\tilde{S}_r(r)=\frac{1}{r}\int_{0}^{r}d\xi\xi\tilde{u}(\xi)=\frac{1}{2}\frac{\partial \tilde{S_z}(r)}{\partial r}\nonumber\\
-\frac{p+2}{2r}\int_{0}^{r}d\xi\xi\tilde{S_z}(\xi)+\frac{1}{2r}\int_{0}^{r}d\xi\xi S_z(\xi,t=0).
\label{LTSrw}
\end{eqnarray}

The boundary conditions (\ref{BCCircle}) are used to find the values of $A_{1,2}$ in Eq. (\ref{LTwSol}). For this purpose, we
Laplace transform Eqs. (\ref{BCCircle}) and employ Eq. (\ref{LTSrw}) to find
\begin{eqnarray}
\frac{\partial^2 \tilde{S_z}(a)}{\partial r^2}+\frac{p+2}{a^2}\int_{0}^{a}dr r\tilde{S_z}(r)-p\tilde{S_z}(a)\nonumber\\
=\frac{1}{a^2}\int_{0}^{a}dr r S_z(r,t=0)-S_z(a,t=0),\;\;
\label{BCw1}\\
\frac{\partial \tilde{S_z}(a)}{\partial r}+\frac{p+2}{a}\int_{0}^{a}dr r\tilde{S_z}(r)
=\frac{1}{a}\int_{0}^{a}dr r S_z(r,t=0).\;\;
\label{BCw2}
\end{eqnarray}
$A_{1,2}$ are obtained from Eqs. (\ref{BCw1}) and (\ref{BCw2}) complemented by Eq. (\ref{LTwSol}) and the initial conditions (\ref{ICS}).
Finally, the Laplace transform of $z$-component of spin polarization is written as
\begin{eqnarray}
\tilde{S_z}(r)=\frac{2S_0}{(p+2)D(p)}
\left[(p+2-k_{2}^2)\frac{J_1(k_{2}a)}{k_{2}}J_0(k_{1}r)\right.\nonumber\\
\left.-(p+2-k_{1}^2)\frac{J_1(k_{1}a)}{k_{1}}J_0(k_{2}r)\right]
+\frac{S_0}{p+2},
\label{LTw}
\end{eqnarray}
where the following notation is used:
\begin{eqnarray}
D(p)=2(p+2)(k_{2}^2-k_{1}^2)\frac{J_1(k_{1}a)J_1(k_{2}a)}{ak_{1}k_{2}}\nonumber\\
+[(p+2)(k_{1}^2-1)-pk_{2}^2]J_0(k_{1}a)\frac{J_1(k_{2}a)}{k_{2}}\nonumber\\
-[(p+2)(k_{2}^2-1)-pk_{1}^2]J_0(k_{2}a)\frac{J_1(k_{1}a)}{k_{1}}.
\label{Det}
\end{eqnarray}
The Laplace transform of $r$-component of spin polarization is found combining Eqs. (\ref{LTSrw}), (\ref{LTw}) and (\ref{ICS}):
\begin{eqnarray}
\tilde{S}_r(r)=\frac{S_0}{(p+2)D(p)}
\left[(k_{1}^2+p+2)(k_{2}^2-p-2)\frac{J_1(k_{2}a)}{k_{2}}\right.\nonumber\\
\left.\times\frac{J_1(k_{1}r)}{k_{1}}-(k_{2}^2+p+2)(k_{1}^2-p-2)\frac{J_1(k_{1}a)}{k_{1}}\frac{J_1(k_{2}r)}{k_{2}}\right]. \;\;
\label{LTSr}
\end{eqnarray}

\begin{figure}[t]
 \begin{center}
\includegraphics[angle=0,width=8.0cm]{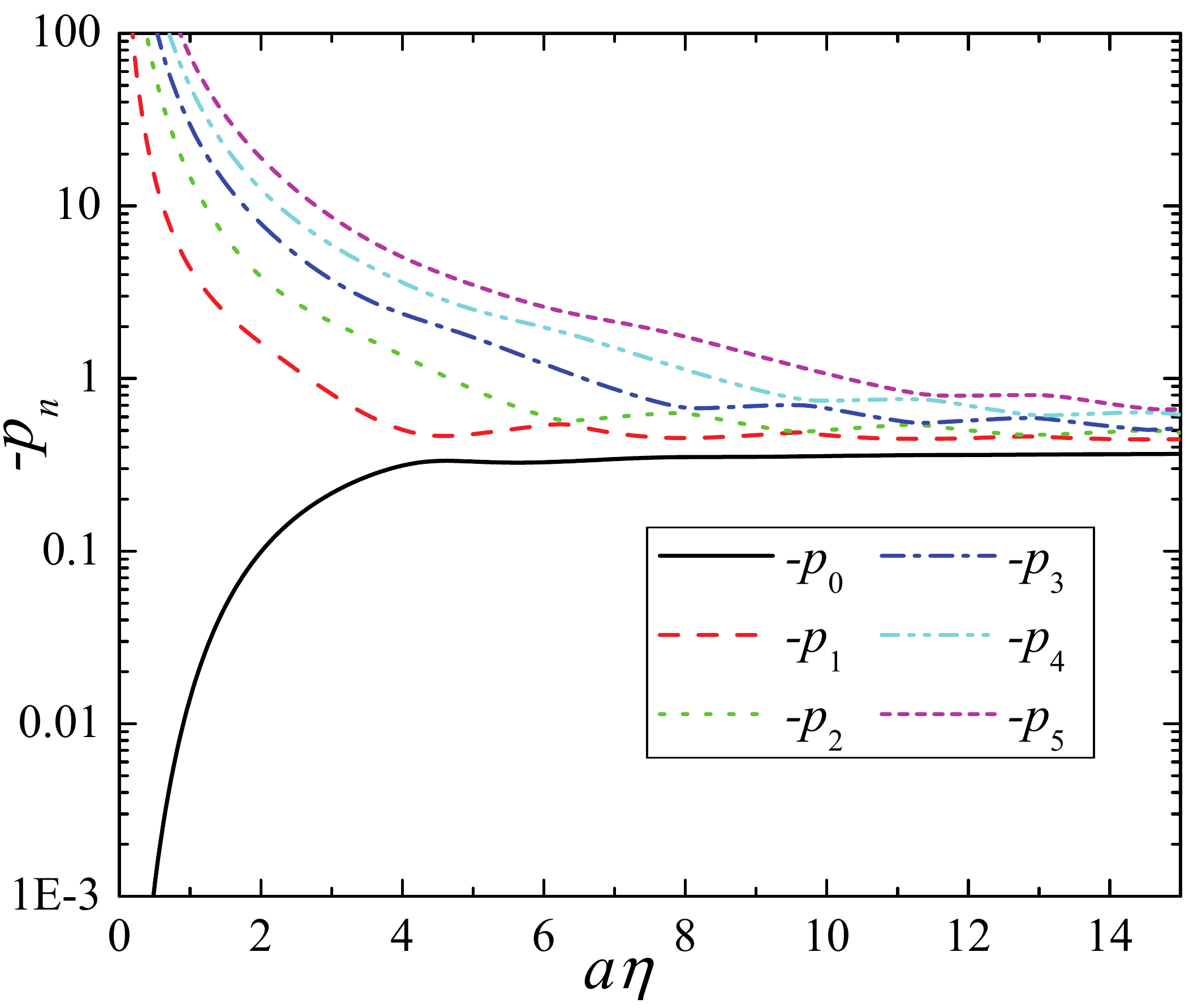}
\end{center}
 \caption{(Color online) First six poles of $\tilde{S_z}(r)$ and $\tilde{S}_r(r)$ versus the ring radius. These poles are proportional to the slowest six spin relaxation rates. \label{fig1}}
\end{figure}

The inverse Laplace transform of Eqs. (\ref{LTw}) and (\ref{LTSr}) provides the time-domain components of spin polarization.
It is important to note that the right-hand sides of Eqs. (\ref{LTw}) and (\ref{LTSr}) do not change under a permutation of $k_1$ and $k_2$. This means that these functions are one-valued functions in the whole complex plane of $p$ despite the square root in Eq. (\ref{kpm}). As a result, $\tilde{S_z}(r)$ and $\tilde{S}_r(r)$ are meromorphic functions. The poles of these functions are defined by
the equation
$D(p)=0$, which have an infinite number of roots $p_n$, $|p_n|\rightarrow \infty$ as $n\rightarrow \infty$. All poles are characterized by $\textnormal{Im}(p_n)=0$ and $\textnormal{Re}(p_n)<0$. Note that, generally, $p=-2$ is not a pole of both $\tilde{S_z}(r)$ and $\tilde{S}_r(r)$. Referring to Fig. \ref{fig1}, the positions of poles depend on the circle radius. At small values of $a\eta$, the values of all $|p_n|$ $(n=1,2,...)$  increase with decrease of $a \eta$ except of $|p_0|$ whose value decreases. Basically, this is the most interesting pole describing the asymptotic spin relaxation at long times. $p_0$  is always located between $-7/16$ and $0$ and tends to zero when $a\eta \rightarrow 0$, and to $-7/16$ when $a\eta \rightarrow \infty$.
In the limit of small $a$, an analytical expression for $p_0$ can be found. Expanding Eq. (\ref{Det}) over small $p$ and $a$, we get (using dimensional units for $a$)
\begin{equation}
-p_0=\frac{1}{48}\left(a\eta \right)^4-\frac{7}{768}\left(a\eta \right)^6+ O \left( \left(  a\eta \right)^8\right) . \label{p0}
\end{equation}
This expression is basically valid when $a\eta <1$. We also note that the first term in the right-hand side of Eq. (\ref{p0}) was previously reported in Ref. \onlinecite{Lyubinskiy06a}. It can also be shown that for $n=1,2,...$,  $|p_n|\sim C_n (a\eta)^{-2}$ when $a\eta\ll 1$.

\begin{figure}[t]
 \begin{center}
\includegraphics[angle=0,width=8.0cm]{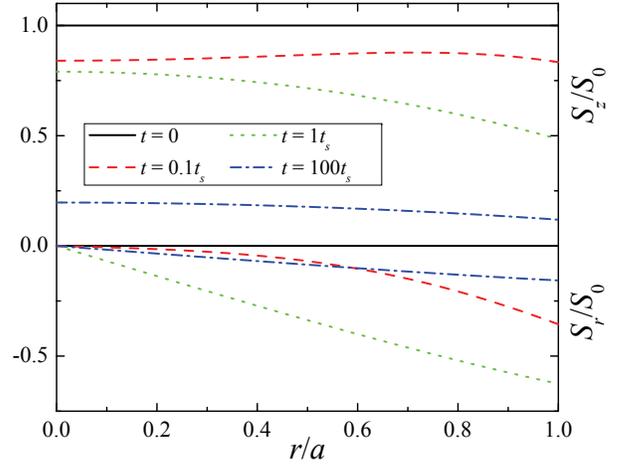}
\end{center}
 \caption{(Color online) Radial distributions of $S_r$ and $S_z$ components of spin polarization at different moments of time.  This plot was obtained at $a\eta=1$.  \label{fig3}}
\end{figure}

\begin{figure}[t]
 \begin{center}
\includegraphics[angle=0,width=8.0cm]{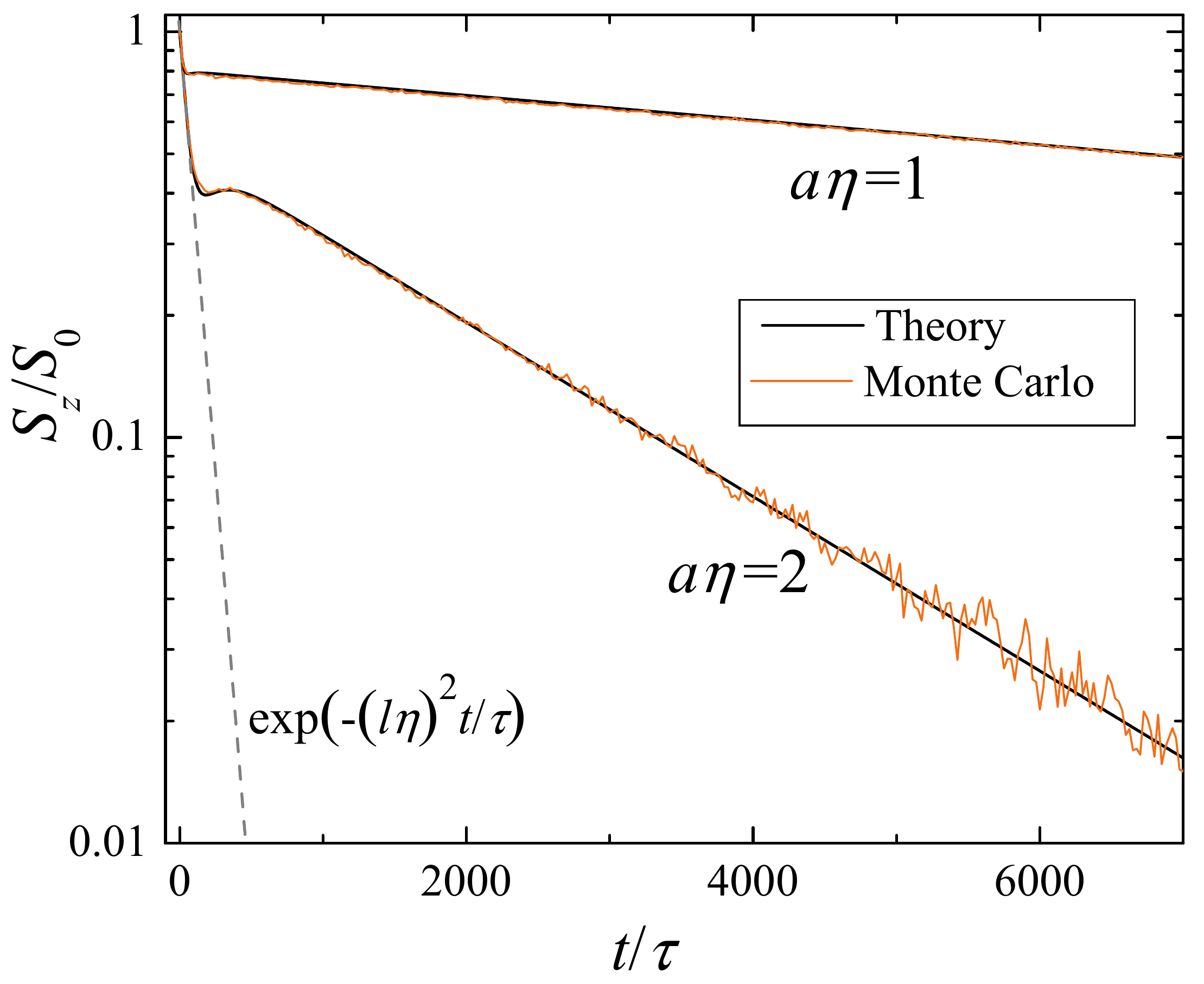}
\end{center}
 \caption{(Color online) Time dependence of $S_z$ component of spin polarization at $r=0$ obtained using Eq. (\ref{w}) and Monte Carlo calculations. The dashed line corresponds to the usual D'yakonov-Perel' relaxation in infinite 2D systems. This plot was obtained using the parameter value $\eta l=0.1$. \label{fig2}}
\end{figure}

The time-domain components of spin polarization are written using the normal dimensional units of time and coordinates as
\begin{eqnarray}
S_z(r,t)=\sum_{n=0}^{+\infty}\frac{2S_0 e^{p_n D\eta^2 t}}{(p_n+2)D'(p_n)}
\left[(p_n+2-k_{2n}^2)\frac{J_1(k_{2n}a\eta)}{k_{2n}}\right.\nonumber\\
\left. \times J_0(k_{1n}r\eta)
-(p_n+2-k_{1n}^2)\frac{J_1(k_{1n}a\eta)}{k_{1n}}J_0(k_{2n}r\eta)\right] \;\;
\label{w}
\end{eqnarray}
and
\begin{eqnarray}
S_r(r,t)=\sum_{n=0}^{+\infty}\frac{S_0 e^{p_n D\eta^2 t}}{(p_n+2)D'(p_n)}
\left[(k_{1n}^2+p_n+2)\right.\nonumber\\
\left.\times(k_{2n}^2-p_n-2)\frac{J_1(k_{2n}a\eta)}{k_{2n}}\frac{J_1(k_{1n}r\eta)}{k_{1n}}
\right.\nonumber\\
\left.-(k_{2n}^2+p_n+2)(k_{1n}^2-p_n-2)\frac{J_1(k_{1n}a\eta)}{k_{1n}}\frac{J_1(k_{2n}r\eta)}{k_{2n}}\right]. \;\;\;
\label{Sr}
\end{eqnarray}
Eqs. (\ref{w}) and (\ref{Sr}) represent the main result of this work describing the time-dependence of spin relaxation in the circle. Basically, three main stages of spin relaxation in small systems can be identified (this separation is appropriate at $a \eta \lesssim 3$, when, as it follows from Fig. \ref{fig1}, $p_0$ is well separated from all other poles). The first (initial) stage of spin relaxation takes place at $t \lesssim a^2/(16D)$, where $a^2/(4D)$ is the time it takes for an electron to diffusively propagate over a distance $a$. During this stage, the most of electrons in the circle's center still do not 'know about' the presence of the boundary and, therefore, the spin relaxation occurs essentially as in the bulk (accordingly to the standard 2D D'yakonov-Perel' spin relaxation theory). In the second stage of spin relaxation, when $a^2/(16D) \lesssim t \lesssim 4/(|p_1| D\eta^2)$, several exponentially decaying terms play the main role in Eqs. (\ref{w}) and (\ref{Sr}). During this stage, a slow-decaying spin polarization profile establishes. In such a profile, $|S_r|$ increases with $r$ as we demonstrate in Fig. \ref{fig3}. The last third stage of spin relaxation is a slow single-exponent decay of the slow-decaying spin polarization profile established during the second stage. This process occurs at long times, namely, when $4/(|p_1| D\eta^2) \lesssim  t$. The analytical expressions describing the shape of the slow-decaying spin polarization profile can be easily inferred from Eqs. (\ref{w}) and (\ref{Sr}) in the long-time limit. All three stages of spin relaxation can be easily distinguished in Fig. \ref{fig2}.

In order to obtain an additional insight on spin relaxation in 2D circle, we have performed extensive Monte Carlo
simulations. All specific details of the Monte Carlo simulations approach can be found in Refs. \cite{Kiselev00a} and \cite{Saikin05a} and will not be repeated here. We just mention that the Monte Carlo simulation program uses a semiclassical description of electron space motion and quantum-mechanical description of spin dynamics. A spin conservation condition was used for electrons elastically scattered from system boundaries. Generally, all obtained Monte Carlo simulation results are in perfect quantitative
agreement with our analytical predictions thus confirming our analytical theory of spin relaxation in finite-size systems.
A comparison of selected analytical and numerical curves is given in Fig. \ref{fig2}.

In summary, we have found an analytical solution for the problem of electron spin relaxation in the circle.
It is shown that a small but non-vanishing spin relaxation exists even in small systems at long times. Consequently, it is not possible to completely eliminate the electron spin relaxation by
reducing the system size, although the relaxation rate is dramatically suppressed in small-size systems. Basically, the spin relaxation process can
be separated into three stages including an initial region of fast bulk-type relaxation, a transition region where the relaxation is described by a combination of several
exponentially decaying functions and a region of slow exponential decay at long times. Our results can be easily verified experimentally.

\bibliography{spin}

\end{document}